# Coexistence of metallicity and superconductivity in adjacent bilayers of a high-$T_C$ superconductor

V. B. Zabolotnyy[1], S. V. Borisenko[1], A. A. Kordyuk[1,2], J. Geck[1], D. Inosov[1], A. Koitzsch[1], J. Fink[1], M. Knupfer[1], B. Büchner[1], S.-L. Drechsler[1], L. Patthey[3], V. Hinkov[4], B. Keimer[4]

[1]*Leibniz-Institute for Solid State Research, IFW Dresden, P.O. Box 270016, D-01171, Dresden, Germany*
[2]*Institute of Metal Physics, 03142 Kyiv, Ukraine*
[3]*Swiss Light Source, Paul Scherrer Institut, CH-5234 Villigen, Switzerland*
[4]*Max-Planck-Institut für Festkörperforschung, 70569 Stuttgart, Germany*

**Experimental studies of the electronic structure remain the basic means for understanding the nature of high-temperature superconductivity (HTSC) and testing relevant theoretical models. Appreciable contributions to establish the overall picture in HTSC have recently been made by investigations on the charge dynamics in BSCCO (ref. 1) and the spin dynamics in YBCO (ref. 2 – 5), using Angle Resolved Photoemission Spectroscopy (ARPES) and Inelastic Neutron Scattering (INS), respectively. Concentration of each of these techniques on a "suitable" compound turns out to be a barrier on the way to a crucial quantitative test allowing to support or discard spin fluctuations (interaction between the charge and spin degrees of freedom) as a possible origin for the pairing in doped cuprates.**

**Here we solve the long-standing puzzle of ARPES on YBCO (ref. 6 – 11) by showing that the photoelectron spectrum of YBCO generally includes two components: One from the topmost anomalously overdoped metallic $CuO_2$ bilayer and the other from the next superconducting bilayer that retains the bulk properties. Our findings clearly show the opening of a large superconducting gap and, for the first time, demonstrate the anisotropic renormalization in the bulk component of YBCO, supporting the universality of these effects for different cuprate families. With our study we re-open this cuprate family for new systematic ARPES investigations.**

Active ARPES research on YBCO-123 all but ended five years ago due to the presence of the controversial "surface state", which was considered to dominate the spectra, masking other bulk related features[6–11]. As a result, the principal issues, such as the presence of the superconducting gap, the band renormalization or the basic bilayer splitting have been left without definite or any answer at all. However, these central questions need to be addressed and clarified before one attempts to learn about the many-body effects responsible for superconductivity. Recently improved experimental techniques have enabled us to revisit this problem.

Fig. 1 shows the experimental Fermi surface (FS) map for near optimal doped untwinned $YBa_2Cu_3O_{6.85}$. In contrast to many previous studies, we use higher excitation energies (50–60 eV vs. 15–30 eV) that enhance the intensity of the Fermi level (FL) crossings of both bonding and antibonding bands and allow examination of a larger area in $k$-space. The visual agreement with the band structure calculations[12] is remarkable. There are two hole-like sheets of the Fermi surface centred around the S point. A set of one-dimensional features running along the Γ-X direction is a clear indication for the presence of states related to the chain substructure of YBCO-123. Even the form of the antibonding sheet of the FS develops W-like undulations close to the Y point. It is also remarkable, that unlike Bi-2212 (ref. 13), the bilayer splitting along the FS is more isotropic, so that even along the nodal Γ-S cut its value remains comparable to that along the antinodal Y-S cut. To further substantiate our identification of the bands, we provide several typical intensity plots in the panels (a)-(h) of Fig. 1. Each of these plots shows the photo intensity as a function of binding energy and momentum along selected directions in $k$-space with the bright features tracing band dispersion. Here, in analogy to another bilayer compound Bi-2212, one can follow the typical behaviour of the plane derived bonding and the antibonding bands. The spectra in the panels (e)-(h) show the band dispersion along the $k_y$-axis, where the one-dimensional chain bands are expected. Indeed, the corresponding parabolic-like band remains unchanged in all four cuts, giving rise to the pairs of lines parallel to the $k_x$-axis on the Fermi surface map. In image (e) the chain band is enveloped by the antibonding band and part of the bonding band can be seen in the left bottom corner dispersing beyond the image frame. The images (f)-(g) demonstrate how the $CuO_2$-plane derived bands cross the FL closer and closer to the Γ-X line when moving towards the X point, so that finally the chain band can be found in-between the bonding and antibonding bands. As follows from the spectra the relative position of the chain band with respect to the plane bands is not much different from the one predicted by LDA. Therefore we find a good qualitative agreement between our data and LDA calculations[12] for the electronic structure of YBCO-123.

At this stage it might be difficult to understand the reasons for the problems encountered in previous experimental studies. However, closer data analysis reveals that the issue is complex. First, we find no evidence for the superconducting gap, even though the spectra were measured at T=30K, i.e. deep in the superconducting state. In addition to a lack of the spectral weight shift at the FL, examination of the band dispersions shows that there is no typical BCS-like band bending, which is normally observed when the gap opens. Subsequent measurements done with samples of the same stochiometry proved the consistency of this effect. Further, the ability to cover several Brillouin zones in our measurements has allowed us to establish another surprising fact. Owing to an extensive FS mapping we determined an area of the bonding and antibonding FS sheets (Fig 1) and, consequently, hole doping. Although the nominal doping of the samples was close to optimal, these measurements show that the actual doping level of the CuO$_2$ planes is about 31 %, which is far beyond the superconductivity dome and corresponds to a purely *metallic* region in the generic phase diagram of hole-doped HTSC. Such an anomalous overdoping not only provides a unique opportunity to study previously inaccessible part of the phase diagram of YBCO-123, but also explains the absence of the superconducting gap and related renormalization effects both in these and in previous photoemission experiments.

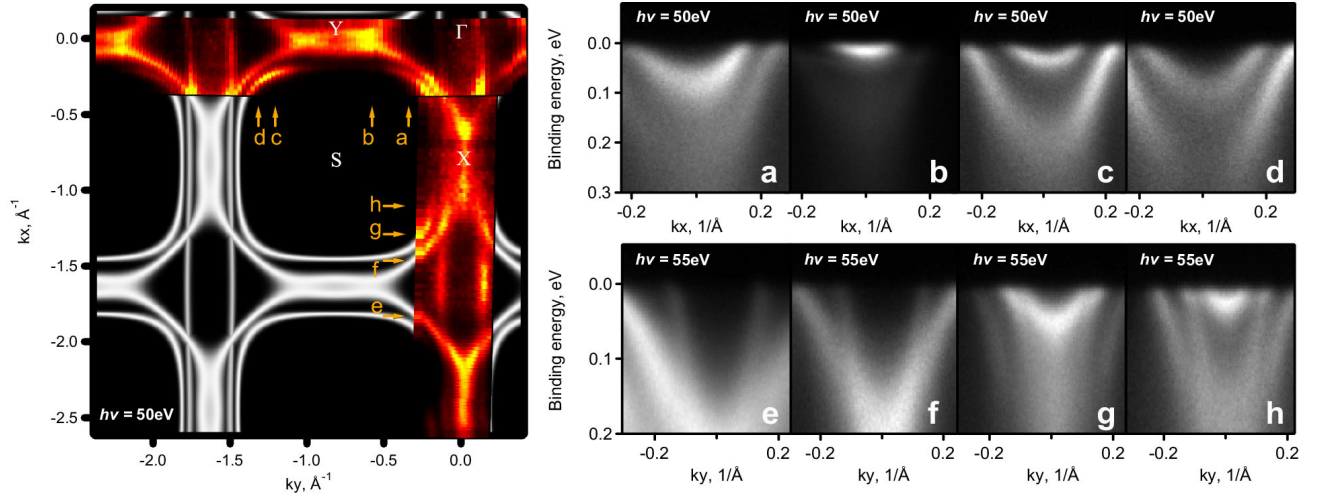

**Figure 1.** Experimental band structure of untwinned YBa$_2$Cu$_3$O$_{6.85}$. The left panel contains two experimental FS maps measured along Γ-X and Γ-Y directions and the tight-biding fit to it. **a-d,** Energy-momentum intensity maps at various $k_x$ values indicated in the FS map. The projection of $k_y$ for these spectra was held constant; its value is denoted by yellow arrows with corresponding lettering on the FS map. **e-f,** Similar measurements, taken after rotating the sample by 90° around Γ point. The image **e** is the sum of spectra measured with circular polarized light of left and right helicity. The other spectra were measured with linear polarized light.

Moreover, it turns out that the YBCO spectra are not only limited to the overdoped component. Changing the photon energy by only 5 eV ($h\nu$=55 eV) transforms the photoemission intensity as found in panel (b) of Fig. 1 into the spectrum presented in Fig. 2 a. An extra component is easily detected in the vicinity of the X and Y points where it looks like a non-dispersing flat feature with a binding energy of about 40 meV. This feature strongly resembles a mode-renormalized and gapped spectrum of Bi-2212, when in the superconducting state the spectral weight is confined in a narrow energy range between the gap and the onset of scattering by a collective mode[14]. This feature gives rise to energy distribution curve consisting of three peaks, which has previously been reported in ref. 11. At that time the metallic component was interpreted differently: Its antibonding band was believed to be a "surface peak", while the term "hump" was used for the bonding band. This feature appears universally in the superconducting state of all YBCO samples with various doping levels that we have studied. Our temperature dependent measurements, in agreement with previous results[11], unambiguously show that the feature virtually disappears above T$_C$ indicating an intimate relationship with superconductivity. We attribute this feature to the antibonding band of the nominally doped bulk component. In Fig. 2 b, using model spectra, we demonstrate how the superposition of superconducting and metallic spectra results in the rather complex structure observed in the experiment (Fig. 2 a). Further studies showed that the ratio between the intensity of the overdoped and nominally doped components in the spectra is very sensitive to a number of experimental parameters, such as the photon energy, polarization, the quality of the exposed surface after cleavage and Ca-substitution. If controlled, these factors can be beneficial for enhancing the contribution of the superconducting component. In Fig. 3 we show spectra from a partially Ca-substituted sample (Ca$_x$Y$_{1-x}$)Ba$_2$Cu$_3$O$_{7-\delta}$, ($x$=0.15±0.03), where the superconducting component dominates. In panels (a)-(e) one can directly observe the momentum dependence of the leading edge gap (LEG) value with its anisotropic *d*-wave like character. The temperature evolution of the gap is given in Fig. 3 f-j and leaves no doubt as to the superconducting nature of these spectra. A very important observation here is the occurrence of renormalization anomalies in the band dispersion and their strong momentum and temperature dependence. It is also notable that there are practically no renormalization effects due to the mode coupling for the overdoped component (compare to Fig. 1a–h, image (g) contains a visible admixture of the antibonding band of the superconducting component), which demonstrates a substantial doping dependence of the renormalization. It is these dependences that were used to argue that the reason for the unusual renormalization effect in Bi-2212 is the coupling to magnetic excitations[14-15]. With our present data we show that the conclusions about the mode origin made earlier for Bi-2212 are equally applicable for the YBCO case, thereby backing their generality.

In the following we address a possible microscopic interpretation for the observed coexistence of metallicity and superconductivity in YBCO. The absence of an easy cleavage plane in YBCO-123 is a known problem[16] that prevents this systems use in extensive studies by tunneling spectroscopy. Existing data[17–18], however, indicate that the most preferable cleavage bond is between Cu-O and Ba-O layers, leading to two types of the surface termination (Fig. 2c). Since for the YBCO-123 it is the amount of oxygen O(1) (ref. 8) in the chains that drives the doping level of the neighbouring $CuO_2$ bilayers, such a cleavage would destroy the chains and certainly change the doping level of the adjacent bilayers. From this it becomes clear that the easily observed intense anomalously overdoped component in the spectra arises from the topmost bilayer with one particular surface termination that results in its overdoping. The superconducting component, on the other hand, originates primarily from the deeper lying bilayer and can only be observed if special conditions for the exciting radiation and polarization are satisfied. The other type of surface termination should lead to underdoping of the topmost bilayer, which follows from simple charge conservation consideration. To make a rough estimate of its doping level we assume that the average doping of the bilayers adjacent to the cleavage plane is equal to that of the bulk one. Estimating the hole concentration per plane for the bulk by $T_C$ ($x \sim 0.15$) (ref. 19) and the hole concentration for the overdoped component from the FS area ($x \sim 0.31$) we find that the underdoped component has to be practically undoped. Hence it does not provide a contribution to the spectra in agreement with the experimental results.

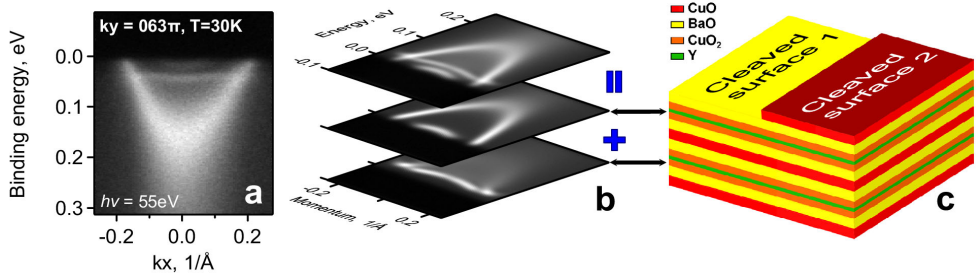

**Figure 2.** Coexistence of superconducting and metallic components. **a**, Experimental spectrum comprising the contribution from normal and superconducting states for $YBa_2Cu_3O_{6.85}$, $k_y$=0.63 $\pi/a$, T=30K, $h\nu$=55eV. **b**, Model image, to explain the spectrum as a superposition of the signal from superconducting and overdoped (metallic) bilayers. The bottommost image is the spectrum of superconducting phase, middle — metallic, and the topmost image is their sum. **c**, Schematic structure of the cleaved crystal. The doping of the $CuO_2$ bilayer nearest to the surface is modified due to disturbed chains (CuO and BaO layers), while the next bilayer remains unimpaired and retains superconductivity.

To conclude we have shown that the main stumbling-block on the way to understand spectra of YBCO-123 was the presence of two components — a superconducting and a metallic one. Our data is consistent with the model where this two component coexists in two neighbouring bilayers at the surface. We show that the superconducting component in YBCO possesses all the necessary attributes like in other HTSC, namely, an anisotropic superconducting gap and band renormalization effects that depend on momentum and doping level and vanish at $T_C$, therefore we support the generality of the mentioned effects. Having understood the YBCO spectra, the next promising step is to use ARPES data and INS results obtained for the same compound to make a quantitative check for consistency of theories aiming to explain high-$T_C$ superconductivity.

**Methods**

The data was collected at the SIS beam line at the Swiss Light Source using a SCIENTA SES 100 electron analyser. Samples were mounted on a high precision cryomanipulator and cleaved in-situ in an ultra high vacuum chamber with a base pressure of $5 \cdot 10^{-11}$ mBar. All spectra, except for the map, were measured with an energy resolution of 12 meV and an angular resolution of 0.2°. For mapping purposes the energy resolution was set to 20 meV. The excitation energies and temperatures are stated within the corresponding spectra or in the caption text. The Fermi level was calibrated using the cold Fermi edge of a fresh silver film evaporated on the sample at the end of experiment. Further details on experimental conditions and sample preparation can be found in ref. 20.

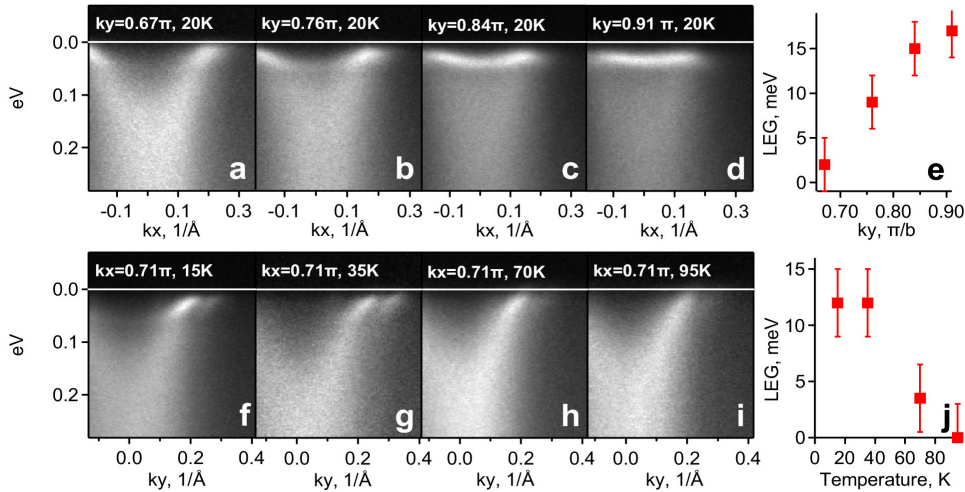

**Figure 3**. Superconducting component. **a-d,** opening of a superconducting gap and enhancement of the band renormalization effects when approaching the antinodal point. **e,** extracted value of the superconducting gap. Evolution of the spectral weight (**f-g**) and the gap (**j**) as a function of temperature. The spectra were measured from a freshly cleaved $(Ca_xY_{1-x})Ba_2Cu_3O_{7-\delta}$, ($x$=0.15±0.03, $T_C$=77 K) using linearly polarized light, $h\nu$ = 60 eV.

High-quality single crystals were synthesized by the solution-growth technique and annealed to the desired oxygen content. The nearly optimally doped sample $YBa_2Cu_3O_{6.85}$ was detwinned by uniaxial mechanical stress at elevated temperatures[5]. The critical temperatures of the samples were measured using the AC susceptibility technique. The critical temperatures were 90 K and 77 K for the pure and Ca substituted samples.

The determination of the doping level of the ovedoped component was based on Luttinger's theorem[22]. The experimental dispersions of the bonding and antibonding bands, where possible, were extracted from the momentum-energy intensity images comprising the FS map shown in Fig. 1. Such a procedure allowed us to locate the FS crossing by the bands with the highest possible accuracy. Finally, the obtained set of FS crossings was fitted with a tight binding model[12] and the area of the FS sheets was calculated. Since we were interested in the doping of the $CuO_2$ bilayers the area of the chain derived band was not taken into account. We also did not account for the area of the small pocket at the S point of the BZ predicted by LDA, as we find no evidence for its existence. According to LDA the area of this pocket may only amount to about $2S_{Pocket}/S_{BZ} \approx 3\%$, which does not influence any of the conclusions of the study.

### Acknowledgments

The project is part of the Forschergruppe FOR538 and was supported by the DFG under Grant No. KN393/4 and by BMBF under Grant No. 05KS40D218. This work was performed at Swiss Light Source, Paul Scherrer, Villigen, Switzerland. We thank R. Hübel for technical support and M. Rümmeli for careful reading of the manuscript.